\documentclass[11pt]{article}

\usepackage{soul}
\usepackage[utf8]{inputenc}
\usepackage{amsmath, amssymb, amsthm, dsfont}

\newtheorem{proposition}{Proposition}

\newtheorem{lemma}{Lemma}
\usepackage{hyperref}
\hypersetup{
    colorlinks=true,
    linkcolor=blue,
    filecolor=magenta,      
    citecolor={blue!90!black},
    urlcolor=cyan,
    pdftitle={Overleaf Example},
    pdfpagemode=FullScreen,
    }

\usepackage[ruled,vlined]{algorithm2e}

\usepackage{graphicx}
\usepackage{dcolumn}
\usepackage{bm}
\usepackage{nicefrac}
\usepackage{empheq}
\usepackage[procnames]{listings}

\usepackage[space=true]{accsupp}
\newcommand{\copyablespace}{\BeginAccSupp{method=hex,unicode,ActualText=00A0}\hphantom{x}\EndAccSupp{}}

\lstdefinelanguage{Julia}%
  {morekeywords={abstract,break,case,catch,const,continue,do,else,elseif,%
      end,export,false,for,function,immutable,import,importall,if,in,%
      macro,module,otherwise,quote,return,switch,true,try,type,typealias,%
      using,while},%
   sensitive=true,%
   alsoother={$},%
   morecomment=[l]\#,%
   morecomment=[n]{\#=}{=\#},%
   morestring=[s]{"}{"},%
   morestring=[m]{'}{'},%
}[keywords,comments,strings]%

\usepackage[utf8]{inputenc}
\usepackage[T1]{fontenc}
\usepackage{mathptmx}
\usepackage{etoolbox}
\usepackage[dvipsnames]{xcolor}
\usepackage{orcidlink}

\usepackage{titlesec}
\makeatletter
\def\@email#1#2{%
 \endgroup
 \patchcmd{\titleblock@produce}
  {\frontmatter@RRAPformat}
  {\frontmatter@RRAPformat{\produce@RRAP{*#1\href{mailto:#2}{#2}}}\frontmatter@RRAPformat}
  {}{}
}%
\makeatother

\newcommand{\thetitle}{
Tensor-SqRA: Modeling the Transition Rates of Interacting Molecular Systems in terms of Potential Energies
}

\newcommand{\theabstract}{
Estimating the rate of rare conformational changes in molecular systems is one of the goals of Molecular Dynamics simulations. In the past decades, a lot of progress has been done in data-based approaches towards this problem. 
In contrast, model-based methods such as the Square Root Approximation (SqRA), directly derive these quantities from the potential energy functions.
In this article we demonstrate how the SqRA formalism naturally blends with the tensor structure obtained by coupling multiple systems, resulting in the tensor-based Square Root Approximation (tSqRA).
It enables efficient treatment of high-dimensional systems using the SqRA and provides an algebraic expression of the impact of coupling energies between molecular subsystems.
Based on the tSqRA, we also develop the Projected Rate Estimation (PRE), a hybrid data-model-based algorithm that efficiently estimates the slowest rates for coupled systems.
In addition, we investigate the possibility of integrating low-rank approximations within this framework to maximize the potential of the tSqRA.
}

\newcommand{\theacknowledge}{This research has been funded by the Deutsche Forschungsgemeinschaft (DFG, German Research Foundation) through the Cluster of Excellence MATH+, project AA1-15 ``Math-powered drug-design'', as well as by the project B05 ``Origin of scaling cascades in protein dynamics'' of the Collaborative Research Center CRC 1114 ``Scaling Cascades in Complex Systems'', DFG Project Number 235221301.}

\usepackage[a4paper, margin=1in]{geometry}
\usepackage[backend=biber,style=numeric-comp, sorting=none]{biblatex}
\usepackage{orcidlink}
\usepackage{authblk}

\makeatletter
\newcommand{\printfnsymbol}[1]{%
  \textsuperscript{\@fnsymbol{#1}}%
}
\makeatother

\author[a,b]{Alexander Sikorski\,\orcidlink{0000-0001-9051-650X}\thanks{These authors contributed equally to this work.}}
\author[c]{Amir Niknejad}
\author[b]{Marcus Weber\,\orcidlink{0000-0003-3939-410X}}
\author[a,b]{Luca Donati\,\orcidlink{0000-0002-6014-6875}\printfnsymbol{1}}

\affil[a]{Freie Universit\" at Berlin}
\affil[b]{Zuse-Institute Berlin}
\affil[c]{College of Mount Saint Vincent}

\date{\today}

\begin{document}

\title{\thetitle} 
\maketitle

\begin{abstract} 
\theabstract 
\end{abstract}

\newcommand{\Qo}{Q^{\mathrm{out}}}
\newcommand{\Qot}{\widetilde{Q}^{\mathrm{out}}}
\newcommand{\Qt}{\widetilde{Q}}
\newcommand{\inv}{^{-1}}
\newcommand{\kron}{\otimes}
\newcommand*\widefbox[1]{\fbox{\hspace{2em}#1\hspace{2em}}}
\newcommand{\One}{\mathds{1}}
\newcommand{\kronsum}{\bigoplus}
\newcommand{\kronprod}{\bigotimes}
\newcommand{\cind}{_C}
\newcommand{\Vi}{\widetilde V} 
\newcommand{\wt}{\widetilde}

\section{Introduction}

Classical molecular systems are modeled by a function $\widetilde{V}(x)$ which provides the potential energy of the system as a function of the $N$-dimensional coordinate vector $x$ of the atoms of the system.
The potential energy function $\widetilde{V}$ is calculated as a sum of interaction energies, and each summand depends on a small subset of the coordinates only.
If we assume that parts of the molecular system are spatially far enough apart from each other, then these parts of the system move almost independently. 
According to this assumption, let the potential energy of the system, e.g.,  be written as 
\begin{equation}
\begin{aligned}
    \widetilde{V}(x):=\widetilde{V}(x_1,x_2) &= V_1(x_1) + V_2(x_2)  + V_c(x_1,x_2)\\
    &= V(x_1,x_2)  + V_c(x_1,x_2)
    \label{eq:Vintro}
\end{aligned}
\end{equation}
where $x_1$ is a $d$-dimensional vector and $x_2$ is an $(N-d)$-dimensional vector of disjoint subsets of coordinates of the entire $N$-dimensional system. 
The potential defined in eq.~\eqref{eq:Vintro} can be investigated from three different points of view: 
(i) $V_1$ and $V_2$ are analyzed individually as \emph{isolated subsystems};
(ii) $V_1$ and $V_2$ are analyzed as a non-interacting \emph{combined system};
(iii) $V_1$ and $V_2$ interact by means of the potential $V_c$ giving rise to a \emph{coupled system}.
In real applications, it is interesting to understand how the term $V_c$ of the coupled system changes the rate of rare transitions of the combined system. 

In order to answer this fundamental question, the typical approach (known as Markov State Modelling\cite{schuette14}) is to produce classical Molecular Dynamics (MD) simulations at the atomistic level, then the simulation data are used to construct transition probability matrices, whose spectral analysis allows the determination of the time scales of the ``macroscopic movements''.

As an alternative to the data-based approach of Markov State Modelling, which is widely known and applied, the model-based Square Root Approximation (SqRA) \cite{Lie13, Donati2018b, Donati2021, Donati2022b} can directly derive transition rate matrices from the potential energy function of the molecular system without taking the detour of generating molecular simulation data.

In this article, we review the fundamentals of SqRA and show how its algebraic structure can be leveraged to calculate the rare event rates of molecular systems with potentials defined as in eq.~\ref{eq:Vintro}.
In particular, we show how SqRA allows to directly represent the kinetic properties of the non-interacting system $V(x_1,x_2)$ in terms of the rate matrices of the isolated subsystems $V_1(x_1)$ and $V_2(x_2)$ using the Kronecker formalism \cite{Loan2000}, thereby circumventing the curse of dimensionality.
Clearly, this simple decomposition breaks down when introducing the coupling term $V_c(x_1,x_2)$.

To alleviate this problem, we developed a tensor formulation of SqRA (tSqRA) which allows to represent the coupling terms by Hadamard products. Since these act on the interacting particles only this enables to inherit as much of the underlying decoupled structure as possible. This reduces the complexity while still providing exact results, avoiding the (computationally expensive) production of molecular simulation data to obtain transition rate matrices of the entire molecular system.
Using tensor algebra to analyze coupled Markov processes is not a new idea, there exist introductory texts to this kind of approaches, see Dayar~\cite{Dayar2012} or  Ludyk~\cite{Ludyk2018}. 
Also analytical and linear algebraic methods are well-established and widely known, see Pollock~\cite{Pollock2011} and Jokar~\cite{Jokar2009}. 
Not only in theory, but also in applications, these methods have been used intensively, see Fernandez~\cite{Fernandez2016} and Ching~\cite{Ching2008}.
However, our manuscript goes one step further than the existing theory and methods by combining the algebraic form of SqRA which provides an explicit link between transition rates and potential energy given by with the tensor algebra approach.

Furthermore, we introduce the Projected Rate Estimation (PRE), a hybrid data-model-based method to obtain the global transition rates of coupled systems as refinements of the transition rates of non-interacting subsystems by carrying out only a few local simulations.

The presented methods do not exploit any low-rank structures in the problem yet and therefore cannot be applied directly to larger systems. 
However, we demonstrate how the provided formalism lends itself to future low-rank approximations via tensor-trains or -networks, thereby facilitating computations on larger scales \cite{Ramanathan2004, Khoromskaia2015, Lucke2021, gelss2017tensor}.

\section{Theory}
Originally, SqRA \cite{Lie13, Kube07} was invented to solve the problem of finding a transition rate matrix $Q \in \mathbb{R}^{M\times M}$ of a molecular system with an Euclidean state space discretized into $M$ subsets, such that this matrix is reversible and has a predefined stationary distribution $\pi\in \mathbb{R}^M$. 
It turned out that a regular grid discretization of the state space guarantees a simple algebraic form of the matrix $Q$ and that increasing the size of the grid $M\rightarrow \infty$ leads to a convergence of $Q$ towards the Fokker-Planck operator of the underlying overdamped Langevin dynamics of the system \cite{Heida2018}. 
Note that for the application of SqRA, a regular grid can be constructed by a hypercubic grid which can be seen as a Voronoi tessellation with equidistant center points in each direction.
In many applications of SqRA, it is assumed that the discretization of the state space is fine enough to approximate the continuous stationary distribution by the pointwise evaluation of the Boltzmann distribution at the centers of the cells. 
In the following sections we will show how the algebraic structure of the SqRA lends itself to the application to combined and even coupled systems.

In section \ref{sec:theoryA} we start by introducing the necessary foundations of the SqRA. Section \ref{sec:theorycombined} then shows how the SqRA matrices for the combined system can be composed from the SqRA matrices of the isolated systems using Kronecker products and sums.
In section \ref{sec:theorycoupling} we show how these matrices are changed by the introduction of a coupling between the subsystems.
Section \ref{sec:theorytensor} provides the most general formulation in terms of tensors and Hadamard products paving the way for future low-rank developments.
\subsection{Linear Algebraic Form of SqRA}\label{sec:theoryA}
Consider an $N$-dimensional dynamical system governed by the potential energy function $V(x):\mathbb{R^N} \rightarrow \mathbb{R}$, where the state space $x$ is discretized by a hypercubic grid with $M$ center points $x_i$ with $i=1,2,\ldots, M$. 
The stationary distribution $\pi$ is given by the Boltzmann distribution with entries
\begin{eqnarray}
    \pi_i := \pi(x_i) = \exp\left(-\frac{1}{k_B T} V(x_i)\right)
    \label{eq:StatDistr}
\end{eqnarray} 
where $k_B$ is the Boltzmann constant and $T$ the temperature. 
The SqRA the defines the sparse rate matrix $Q\in \mathbb{R}^{M \times M}$ as 
\begin{eqnarray}
Q_{ij} &=& 
\begin{cases}
\phantom{-}\displaystyle\sqrt{\nicefrac{\pi_j}{\pi_i}}
&\mbox{if  $x_i$ is adjacent to $x_j$}  \\
-
\sum_{\substack{i =1\\ i \ne j}}^M Q_{ij}  &\mbox{if } x_i = x_j  \\
\hspace{20pt} 0 &\mbox{otherwise} 
\end{cases} 
\label{eq:RateMatrix}
\end{eqnarray}
or alternatively in matrix form
\begin{eqnarray}
Q 
&:=& D^{-1} A D - \mathrm{diag}(D^{-1}AD\,e) \cr
&=& D^{-1} A D - E  \cr
&=& \Qo - E \, .
\label{eq:SqRA2}
\end{eqnarray}
Here $D$ is a diagonal matrix with the square roots of the vector $\pi$ on the diagonal, i.e. $D=\mathrm{diag}(\sqrt{\pi})$, $A$ is the the adjacency matrix of the grid, and $e_M=(1,1,\ldots,1)^T$ is a column vector with $M$ entries.
We denote the off-diagonal part of $Q$ by $\Qo = D^{-1} A D$ and the diagonal part $E = \mathrm{diag}(D^{-1}AD\,e_M)$ is chosen in such a way that $Q$ has row sum zero. 
Note that the stationary distribution in eq.~\eqref{eq:StatDistr} does not need to be normalized since the quotient in the SqRA's formula cancels out the normalization constant.
When definining $Q$ this way it has unit-less entries.
To obtain a proper rate matrix, it is necessary to multiply each entry $Q_{ij}$ by a flux term $\Phi_{ij}$ that depends on the grid geometry and the diffusion \cite{Donati2021}. 
For simplicity we assume a regular grid and constant diffusion such that we can omit $\Phi \propto 1\,\mathrm{[time\ units]^{-1}}$.

With this assumption, the matrix $Q$ is a transition rate matrix. 
The rows of $\Qo$ comprise the outgoing rates from one given state (subset) of the system to the adjacent states and the diagonal entries of $Q$ are negative and represent the ``total exit rate'' from a given state.
Since these are also the entries of $E$, we will also refer to $E$ as the ``exit rate matrix''.
%
%
Note that $Q$ can be seen as a similarity transform of $A-E$:
\begin{eqnarray}
Q 
&=&  D^{-1} (A-E) D \, .
\label{eq:SqRA}
\end{eqnarray}
Thus, eigenvalues of $Q$ and $A-E$ coincide.
As a consequence the implied time-scales of molecular systems, which are derived from these eigenvalues, only depend on the ``total exit rates'' and on the adjacency of the discretized states of the system.

In this regard $A-E$ is like a decomposition of the process into a ``entropical'' part (adjacency matrix $A$) and a ``energetic'' part (exit rates matrix $E$).
The eigenvectors (or their sign structure) of $Q$ on the other hand can be used to identify macro-states in the form of conformations or metastabilities by algorithms like PCCA \cite{Deuflhard00} or PCCA+ \cite{Deuflhard05}.
If $v$ is an eigenvector of $A-E$, then $D^{-1}v$ is the corresponding eigenvector of $Q$ (sharing the same sign structure with $v$).

Describing transitions in terms of rate matrices is only one possible notation.  A very common way of analyzing Markov processes is the use of transition probability matrices instead of rate matrices. Our approach can be transfered to the framework of transition matrices.
A Markov Chain which is observed for a specific lag time $\tau$ (the time unit is defined by the flux in eq.~\eqref{eq:SqRA}) is represented by a conditional probability matrix
\begin{equation}
    K^\tau = \exp(\tau Q)\,,
    \label{eq:Koopman}
\end{equation}
where $\exp(\cdot)$ denotes the matrix exponential.
Eq.~\eqref{eq:Koopman} defines the discrete version of the Koopman operator that transports observable functions $f\in L^{\infty}$ forward in time.
Correspondingly, its transpose is a discretization of the propagator $P^\tau$ which transports probability densities $\rho\in L^{1}$.
The entries of the $i$th row of $K^\tau$ quantify the conditional probability for a system that starts in this state $i$ to end up in the respective states in the time span $\tau$. 
In contrast to $Q$, the matrix $K^\tau$ is usually a dense matrix. 
However, $K^\tau$ has the same eigenvectors like $Q$ and if $\lambda$  is an eigenvalue of $Q$, then $\exp(\tau \lambda)$ is an eigenvalue of $K^\tau$. In this regard many results and methods developed for $K^\tau$ or $P^\tau$ directly carry over to $Q$.
%


\subsection{The Kronecker Formalism of SqRA for Combined Subsystems}
\label{sec:theorycombined}
When considering two isolated systems consisting of $M_1$ and $M_2$ possible states respectively, the combined system can be in one of $M=M_1 \times M_2$ possible states and the corresponding $Q$ matrix is of size $M_1\cdot M_2 \times M_1\cdot M_2$.
However, since they do not interact they should still be described by the two individual system's rates matrices $Q_1, Q_2$ which are of size $M_1 \times M_1$ and $M_2 \times M_2$ only, therefore providing a way more compact representation. 
In this section, we will show how the SqRA allows for such a compact representation in the form of Kronecker products and sums.

We first consider the case of $n=2$ combined subsystems, respectively of dimensions $L_1$ and $L_2$ such that $L_1+L_2=N$, before stating the more general result for arbitrary $n$. 
The combined potential energy is the sum of the subsystems potentials $V_1:\mathbb{R}^{L_1}\rightarrow \mathbb{R}$ and $V_2:\mathbb{R}^{L_2}\rightarrow \mathbb{R}$:
\begin{eqnarray}
V(x) := V(x_1,x_2) = V_1(x_1) + V_2(x_2) \, ,
\label{eq:DecomposedPotential}
\end{eqnarray}
where $x_1\in \mathbb{R}^{L_1}$ and $x_2\in \mathbb{R}^{L_2}$ are the state vectors of the subsystems.
Similarly, due to its exponential form, the overall stationary distribution defined in eq.~\eqref{eq:StatDistr} factors into the product of the marginal distributions of the subsystems
\begin{eqnarray}
    \pi(x)  
    :=
    \pi(x_1,x_2)
    = 
    \pi_1(x_1)\pi_2(x_2)
 \, .
\end{eqnarray}

The hyper cubic grid discretizing the Euclidean state space is given by the combination of the grids along the two system's sets of coordinates, each consisting of $M_1$ and $M_2$ cells respectively for a total number of $M=M_1\cdot M_2$ cells.
Let us now see how the matrices $A$, $D$, $E$, $\Qo$ and $Q$ introduced in eq.~\eqref{eq:SqRA2} are rewritten in terms of the smaller matrices associated with the two subsystems.
\subsubsection{Adjacency matrix}
The matrix $A$ of the coupled system is given by the Kronecker sum of the corresponding adjacency matrices of the subsystems:
\begin{eqnarray}
\label{eqn:kronsumA}
A 
&=& A_1\oplus A_2 \cr 
&=& A_1 \kron I_2 + I_1 \kron A_2 \, ,
\label{eq:Adeco}
\end{eqnarray}
where $I_1$ and $I_2$ are respectively two identity matrices of size $(M_1 \times M_1)$ and $(M_2 \times M_2)$.
\subsubsection{Diagonal matrix}
The stationary distributions $\pi_1$ and $\pi_2$ of the subsystems are used to build the diagonal matrices 
$D_1=\mathrm{diag}(\sqrt{\pi}_1)$ 
and 
$D_2=\mathrm{diag}(\sqrt{\pi}_2)$,
and the matrix $D$ is given by Kronecker product of the corresponding subsystems matrices:
\begin{eqnarray}
D &=& D_1 \kron D_2 \,.
\label{eq:Ddeco}
\end{eqnarray}
\subsubsection{Off-diagonal matrix}
Inserting eq.~\eqref{eq:Adeco} and eq.~\eqref{eq:Ddeco} into the off-diagonal part of the SqRA (eq.~\eqref{eq:SqRA2}) we obtain
\begin{eqnarray*}
    \Qo 
    &=& D^{-1}AD  \cr
    &=&(D_1\inv \kron D_2\inv) (A_1 \kron I_2 + I_1 \kron A_2) (D_1 \kron D_2) \cr
    &=&(D_1\inv \kron D_2\inv)(A_1\kron I_2)(D_1\kron D_2) + (D_1\inv \kron D_2\inv)(I_1\kron A_2)(D_1\kron D_2)\cr
    &=&(D_1\inv A_1 \kron D_2\inv I_2)(D_1 \kron D_2)+ (D_1\inv I_1\kron D_2\inv A_2)(D_1\kron D_2)\cr
    &=& D_1\inv A_1 D_1\kron I_2 + I_1 \kron D_2\inv A_2D_2\, ,
\end{eqnarray*}
and we see that the off-diagonal part indeed decomposes into the Kronecker sum of the individual systems
\begin{eqnarray}
    \Qo 
    &=& \Qo_1 \oplus \Qo_2.
    \label{eq:Qodeco}
\end{eqnarray}
\subsubsection{Rate matrix}
Since transition probabilities are given by the product probabilities for the respective transitions of the subsystems, the discretization of the Koopman operator of the full system can be decomposed as \cite{Dayar2012}
\begin{eqnarray}
K^\tau
&=&
K^\tau_1 \kron K_2^\tau \cr
&=& \exp(\tau Q_1)\kron \exp(\tau Q_2) \cr
&=& \exp(\tau (Q_1\oplus Q_2)) \,,
\label{eq:Kdeco}
\end{eqnarray}
where we applied the Kronecker sum rule for exponential matrices in the last line.
This implies
\begin{eqnarray}
Q=Q_1\oplus Q_2 \, .
\label{eq:Qdeco}
\end{eqnarray}
\subsubsection{Exit rate matrix}
Applying eq.~\eqref{eq:Qdeco} and eq.~\eqref{eq:Qodeco} yields
\begin{eqnarray*} 
    Q 
    &=& Q_1 \oplus Q_2 \cr
    &=& (\Qo_1 - E_1) \oplus (\Qo_2 - E_2) \cr
    &=& (\Qo_1 - E_1) \kron I_2 + I_1 \kron (\Qo_2 - E_2) \cr
    &=& \Qo_1 \oplus \Qo_2 - E_1 \oplus E_2\, , \cr 
\end{eqnarray*}
from which one derives  
\begin{eqnarray}
E = E_1\oplus E_2 \, .
\label{eq:Edeco}
\end{eqnarray}

For an arbitrary number $n$ of subsystems we summarize this finding in the following proposition:

\begin{proposition}
\label{prop:combined}
    Consider an $N$-dimensional system with potential $V:\mathbb{R}^N\rightarrow \mathbb{R}$ and stationary distribution $\pi:\mathbb{R}^N\rightarrow \mathbb{R}$ defined on a state space that can be discretized by an $N$-dimensional hypercubic grid.
    Assume that the coordinates can be partitioned into $n$ subsets of  size $L_1, L_2, \dots, L_n$ each, with $\sum_{i=1}^n L_i = N$, such that the potential $V$ is written as
    $$
    V(x) = \sum_{i=1}^n V_i(x_i) \, ,
    $$
    where each potential $V_i:\mathbb{R}^{L_i}\rightarrow \mathbb{R}$ is an $L_i$-dimensional function that depend only on the $i$th subset of coordinates.
    Correspondingly, the stationary distribution is rewritten as
    $$
    \pi(x) = \prod_{i=1}^n \pi_i(x_i) \, .
    $$

    For each subsystem $i$, we define 
    the adjacency matrix $A_i$, 
    the diagonal matrix $D_i=\mathrm{diag}\left(\pi_i^{\nicefrac{1}{2}}\right)$, 
    the off-diagonal rate matrix $\Qo_i = D_i\inv A_i D_i$, 
    the exit rate matrix $E_i = \mathrm{diag}\left(\Qo_i \, e_{L_i}\right)$ 
    as well as the rate matrix $Q_i = \Qo_i - E_i$.    
    Then, the SqRA matrices for the entire system are given in terms of the subsystem matrices 
    
    \begin{empheq}[box=\widefbox]{align*}
        A   &= \bigoplus_{i=1}^n  A_i    \cr
        D   &= \bigotimes_{i=1}^n D_i    \cr
        \Qo &= \bigoplus_{i=1}^n  \Qo_i = D\inv A D \cr
        E   &= \bigoplus_{i=1}^n  E_i    \cr
        Q   &= \bigoplus_{i=1}^n  Q_i = \Qo - E 
        \label{box:box1}
    \end{empheq}
        
\end{proposition}

\subsection{Matrix Representation of the SqRA with a Global Coupling Term}
\label{sec:theorycoupling}

The previous section showed that the structure of the SqRA naturally leads to a low-rank representation for the coupled case. We will now study how the results change when adding a coupling between the systems.

As in the previous section we start with the case of $n=2$ subsystems first.
Given the coupling potential $V_c$ the total potential energy of the coupled system is
\begin{eqnarray}
\widetilde{V}(x_1,x_2) 
&=& V_1(x_1) + V_2(x_2) + V\cind(x_1,x_2) \, .
\label{eq:CouplingPotential}
\end{eqnarray}
The unnormalized stationary distribution is decomposed as

\begin{eqnarray}
    \widetilde{\pi}(x_1,x_2)
    = 
    \pi_1(x_1)\pi_2(x_2)\,
    \pi\cind(x_1,x_2)
 \, ,
\label{eq:CouplingStatDistr}
\end{eqnarray}
where $\pi\cind(x_1,x_2) = \exp\left(-\nicefrac{1}{k_B T}V\cind(x_1,x_2)\right)$.
The adjacency relations are not affected by the coupling term, so the adjacency matrix $A$ is the same as in eq.~\eqref{eq:Adeco}.
On the other hand, each entry of the diagonal matrix $D$ is reweighted by the coupling, leading to
\begin{equation}
    \widetilde{D} = D\cind \, \left( D_1 \kron D_2 \right) 
    \label{eq:Ddeco2}
\end{equation}
where $D\cind = \mathrm{diag}\left(\pi\cind^{\nicefrac{1}{2}}\right)$ is the $(M\times M)$ diagonal matrix built with the stationary distribution of the coupling potential $V\cind$, while $D_1$ and $D_2$ are defined as for the combined system.
Similarly, the inverse satisfies
\begin{eqnarray}
        \widetilde{D}\inv = D\cind\inv \,\left( D_1\inv \kron D_2\inv \right)\,.
    \label{eq:Ddeco2inv}
\end{eqnarray}
Note that, given that $D\cind$ is a diagonal matrix, the calculation in eq.\eqref{eq:Ddeco2} can be interpreted as either a matrix-matrix multiplication or an elementwise multiplication. This section focuses on the matrix formalism, but the elementwise interpretation will play a central role in the next section.

The off-diagonal matrix $\Qot$ of the coupled system is written as
\begin{equation}
\begin{aligned}
    \Qot &= \widetilde D^{-1} A \widetilde D \\
    &= D\cind\inv \,\left( D_1\inv \kron D_2\inv \right) (A_1 \oplus A_2) D\cind \, \left( D_1 \kron D_2 \right) \\
    &= D\cind\inv \left(\Qo_1 \oplus \Qo_2\right) D\cind \\
    &= D\cind\inv \Qo D\cind
\end{aligned}
\end{equation}
and thus also $\widetilde E =  \mathrm{diag}\left(D\cind\inv\Qo D\cind \, e_M \right)$. 
According to eq.~\eqref{eq:SqRA2}, the SqRA rate matrix for the coupled system then reads
\begin{equation}\label{eq:Qcoup}
    \begin{aligned}
        \widetilde{Q} &=  \widetilde{D}\inv A \widetilde{D} - \widetilde{E} \\
        &= D\cind\inv  \Qo D\cind - \widetilde{E} 
    \end{aligned}
\end{equation}

In the general case for $n$ subsystems coupled by a single coupling term, Proposition \ref{prop:combined} results are modified as follows:

\begin{proposition}
    Consider the situation of Proposition 1 but with a global coupling potential $V\cind:\mathbb{R}^n \rightarrow \mathbb{R}$. The potential thus decomposes into 
    \begin{eqnarray}
    \widetilde{V}(x_1,x_2,\dots,x_n)
    &=&
    \sum_{i=1}^n V_i(x_i)
    +
    V\cind(x_1,x_2,\dots,x_n) \, .
\end{eqnarray}
with additional coupling term $V_C:\mathbb{R}^n \rightarrow \mathbb{R}$ and the stationary distribution factorizes as
\begin{eqnarray}
    \widetilde{\pi}(x_1,x_2,\dots,x_n)
    = 
    \prod_{i=1}^n \pi_i(x_i)\,
    \cdot \pi\cind(x_1,x_2,\dots,x_n) \, .
\end{eqnarray}
The SqRA matrices for the coupled system are then given by

\begin{empheq}[box=\widefbox]{align}
    &D\cind = \mathrm{diag}[\sqrt{\pi\cind}] \cr
    %
    %
    &\Qot 
    =
    D\cind\inv \Qo D\cind\cr
    &\widetilde{E} = \mathrm{diag}\left[D\cind\inv \Qo D\cind \, e_M \right]\cr
    &\widetilde{Q} = \Qot   \, 
    - \widetilde{E}
    \label{box:box_pert}
\end{empheq}

\end{proposition}
\subsection{Generalisation to the Tensor Formulation for Arbitrary Interactions}
\label{sec:theorytensor}

Using the tensor formalism, previous results can be easily generalized to the case where the potential $\Vi$ is given by a sum of lower order potentials, i.e. potentials that act only on a subset of coordinates, leading to a flexible decomposition in terms of Hadamard products and paving the way for low-rank tensor computations.

To this end let us consider each coordinate as an individual subsystem, i.e. $n=N$, and introduce the space of tensors of order $N$, $T(N) = \mathbb{R}^{M_1 \times ... \times M_N}$.
Each individual state of the (discretized) system can be understood as a single entry of this tensor, then elements $x\in T(N)$ represent distributions or functions over all states.
We furthermore introduce the symbolic multi-indices $I \in \mathcal{I} \subset \mathcal{P} (\{1,...,N\})$ where $\mathcal{P}$ denotes the power set, i.e. the set of all possible subsets of indices that can appear.
We use these multi-indices in subscript to denote the coordinates upon which the individual lower-order potential contributions $V_I$ depend: 
\begin{equation}
\Vi = \sum_{I\in \mathcal{I}} V_{I}(x_I)\,.
\end{equation}
Further we use Greek superscript letters to denote the individual grid positions of those respective coordinates.
For example, the tensor $V_{1,2}\in\mathbb{R}^{M_2\times M_3}$ of order 2 with components $V_{1,2}^{\alpha \beta} = V_{1,2}(x^{\alpha \beta})$, $\alpha=1,..,M_1, \beta=1,...,M_2$,  holds the evaluation of all potential contributions of the combinations of first and second coordinates at the respective product grid.
To each set of indices $I$ corresponds a tensor $D_I$ of order $|I|$ consisting of the square roots of the stationary distribution (eq.~\eqref{eq:StatDistr}) of the corresponding potential
\begin{equation}
    D_I^{\alpha \beta \gamma...} = \exp \left(-\frac{1}{2} \frac{1}{k_B T} V_I^{\alpha \beta \gamma...}\right) \, .
\end{equation}
This tensor holds all the information about the interaction between the indices $I$.
We combine the individual interactions to the whole interaction tensor $D$ of order $N$ by means of the (widened) \emph{Hadamard product}
\begin{equation}
    (a_{ijk} \odot b_{jkl})^{\alpha \beta \gamma \delta} := a_{ijk}^{\alpha \beta \gamma} \cdot b_{jkl}^{\beta \gamma \delta} \, ,
\end{equation}
where we ``broadcast'' or ``widen'' the elementwise multiplication along all dimensions appearing only on one side. 
The SqRA tensor $D \in T(N)$ then decomposes into the Hadamard factors corresponding to the respective lower order potentials
\begin{equation}
    D_{ijkl...} = D_i \odot D_{j} \odot ... \odot D_{ij} \odot ... \odot D_{ijkl...} \, ,
\end{equation}
or shortly
\begin{equation} \label{eqn:dproduct}
    D = \bigodot_{I\in\mathcal{I}} D_I \, .
\end{equation}
In the previous sections, we introduced the adjacency matrix $A$.
In the case of a 1-dimensional system it consists of a sparse matrix with two off-diagonals. 
For the flattened representation of higher-dimensional systems it is a multibanded-matrix with $2\times N$ bands.
However, the \emph{Kronecker sum} representation defined in eq.~\eqref{eqn:kronsumA}, directly translates 
to a tensor representation with entries
\begin{equation}
    A^{\alpha \beta \gamma..., \alpha' \beta' \gamma'...} = 
    A_1^{\alpha \alpha'} \oplus A_2^{\beta \beta'} \oplus ... = 
    A_1^{\alpha \alpha'} \delta^{\beta\beta'} \delta^{\gamma\gamma'} ... +\delta^{\alpha\alpha'}  A_2^{\beta \beta'}  \delta^{\gamma\gamma'} ... + ... \, ,
\end{equation}
with $\delta$ being the usual Kronecker delta: $\delta^{\alpha\alpha'} = 1$ if $\alpha=\alpha'$ and $0$ otherwise.
In this regard we will think of $A$ as a linear map, mapping tensors of order $N$ to tensors of order $N$.
The action of the SqRA tensor $Q:T(N)\rightarrow T(N)$ on a state $x\in T(N)$ can then be computed
via 

\begin{align}
    Qx &= D^{-1} \odot A(D\odot x) - E \odot x
\end{align}
where $E = D^{-1} A(D) \in T(N)$.

Let us now discuss the practical implications of this formulation in terms of computational complexity. 
Let $M$ denote the size of a state vector in $T(N)$.
The action of the adjacency operator $A$ on a tensor state $x \in T(N)$ can be computed by $2NM$ floating point operations.
The regular grid leads to a banded matrix-representation for flattened states, that allows for a very cache-efficient implementation (c.f. appendix \ref{app:juliacode}).
Since $D \in T(N)$ it requires just as much memory as we need to hold the state $x$ in memory.
Note here that when computing $D$ according to eq.~\eqref{eqn:dproduct} we evaluate the potential functions only on grids up to the order of the interaction.

For example, consider a system of $N$ particles in 1-d space with pairwise interactions and a grid of $m$ cells for each particle. 
Using the tensor representation eq.~\eqref{eqn:dproduct} each of the $N\times(N-1)$ pairwise potentials gets evaluated on a grid of the size $m^2$, resulting in $O(N^2m^2)$ potential evaluations, compared to $O(N^2  m^N)$ evaluations for a naive application on the whole grid. 
Similarly, a system of $n=\nicefrac{N}{3}$ particles in 3 dimensional space with bond (pair-wise) and angle (triplet-wise) interactions, discretized to $m$ cells in each of the $N=3n$ coordinates, requires $O(n^3 m^3)$ evaluations instead of $O(n^3 m^N)$.

Put differently: Let $j$ denote the order of the highest order interactions and assume that the number of these interactions is fixed.
The computational effort for computing the $D$ tensor then scales with $O(m^j)$ compared to $O(m^N)$ for the classical evaluation on the whole grid, i.e. it does not depend on the full system dimension.
Note however, that the number of cells $M=m^N$, still grows exponentially in the dimension.
Therefore when using a dense state representation the application of $D$ scale badly.

However, since each $D_I$ acts only on a few modes, low-rank representations of the state, in conjunction with approximate low-rank computations of the Hadamard product allow for a closed low-rank representation of $Q$.
In cases where the dynamics do indeed permit low-rank representations, as would be expected for weakly interacting systems, this method promises to break the curse of dimensionality (see also the discussion in Sec. \ref{sec:disc_sec}).

Finally, note that for the iterated application, just as with the similarity transform in the matrix case, we have
\begin{equation}
    Q^n(x) = D^{-1} \odot (A-E)^n(D\odot x)\, ,
\end{equation}

Thus, in order to compute the spectrum of $Q$ by an iterative solver we merely need the repeated evaluation of $A - E$, resulting in $O((2N+1)M)$ floating point operations.
Considering that $A$ is inherently low-rank (c.f. eq. \eqref{eqn:kronsumA}), using a low-rank state representation promiseS an exponential speedup.

To summarize, using the tensor SqRA (denoted as tSqRA) allows to alleviate the (explicit) exponential dependence of the computation cost for $Q$ in the system dimension, replacing it by the maximal order of interactions. Even though the scalability is still influenced by the state-size (implying an exponential relationship for dense representation), the tensor formalism should naturally facilitate the integration of low-rank techniques to manage this aspect.

\subsection{Coupling of the eigenvalues and the eigenvectors}

We noted previously that the eigenvalues and eigenvectors of $Q$ are of special interest to understand the slow time-scale dynamics of the process. Let us therefore investigate what we can say about the spectrum of the composed and coupled system.
Since $Q$ is similar to $A-E$, which in turn is symmetric, it follows that all eigenvalues are real numbers. 
Furthermore, the leading eigenvalue of $Q$ matrices is $0$ (with a  constant eigenvector) and all other eigenvalues are negative.

Let $0=\lambda_i^{(1)}>\lambda_i^{(2)}\geq \ldots \geq\lambda_i^{({M_i})}$ denote the eigenvalues with respective eigenfunctions $X_i^{(1)}, \ldots, X_i^{(M_i)}$ of the rate matrix $Q_i$ of the $i$th isolated subsystem.
Defining $\Lambda_i=\mathrm{diag}{(\lambda_i)}$ such that $Q_i X_i = \Lambda_i X_i$, for the combined system $Q$ we have 
\begin{equation}
    Q X =  (\bigoplus_i Q_i) (\bigotimes_i X_i) = (\bigoplus_i \Lambda_i) (\bigotimes_i X_i) = \Lambda X \,,
\end{equation}
where
$\Lambda := \bigoplus_i \Lambda_i $ 
is a diagonal matrix with eigenvalues as diagonal entries, and 
$X := \bigotimes_i X_i$ is a matrix with
the eigenvectors of the combined system.
In terms of individual eigenvalues $\lambda^{(j)}$ this comes down to the eigenvalues being a sum of one of each of the isolated system's eigenvalues,
\begin{eqnarray}
   \lambda^{(j)} = \sum_{i=1}^n \lambda_i^{(index(i))} \, ,
   \label{eq:sum_evals}
\end{eqnarray}

where $index_j(\cdot)$ picks an index for each subsystem (depending on $j$). 
The corresponding eigenvector $X^{(j)}$ is then given by the product of the respective subsystem's eigenvectors:
\begin{eqnarray}
    X^{(j)}=\bigotimes_{i=1}^n X^{index(i)}_{(i)} \, .
    \label{eq:kron_sum_evecs}
\end{eqnarray}
When introducing a coupling term, $V\cind$, the spectrum is perturbed from $\lambda^{(j)}$ to $\widetilde{\lambda}^{(j)}$ as well as from $X^{(j)}$ to $\widetilde{X}^{(j)}$. 

The easiest case is the coupling of two subsystems. 
One can observe a repeating algebraic pattern comparing the construction of the rate matrix $Q$, from the adjacency matrix $A$, with the construction of coupled systems from uncoupled systems. 
In eq. \eqref{eq:SqRA2}, we have a conversion from $A$ which represents transition rates with regard to a constant potential energy function into a matrix $Q$ of exit rates with regard to a non-constant potential $V$. 
Interestingly, this is the same kind of linear algebra like the transition from the positive exit rates of the combined system (matrix $\Qo$) with a constant coupling energy to the rate matrix of the coupled system $\widetilde{Q}$ with a non-constant $V_I$. 
It is given by a similarity transform of a positive rate matrix using a diagonal matrix followed by subtracting the diagonal matrix of row sums.
Although this is a simple linear algebraic correspondence between the matrices, it also shows, that a transition from $Q$ to $\widetilde{Q}$ can change the solution of the eigenproblem  largely, as the transition from a pure adjacency $A$ to a molecular system (by regarding the potential $V$) does. 
However, by the structure of the equation
\begin{equation}
\label{eq:similarity}
\widetilde{Q}= D_{C}\inv (Q - \Delta E) D_{C},
\end{equation}
where $\Delta E = \widetilde{E} - E$ is the difference of ``total leaving rates'', one can see that the eigenvalues of $\widetilde{Q}$ are identical to the eigenvalues of $(Q-\Delta E)$. 
Furthermore, the eigenvectors of $\widetilde Q$ are the eigenvectors of $(Q-\Delta E)$ except for a componentwise rescaling using the diagonal matrix $D_{C}\inv$. This rescaling does not have an effect on the sign structure of the eigenvectors. 
\begin{lemma}
    The matrix $\widetilde Q$ is a similarity transform of $(Q-\nabla E)$, the whole change of the timescales is therefore driven by a perturbation of the diagonal only.
\end{lemma}

In conclusion, for a PCCA-based analysis of the influence of a coupling energy term on the slowest processes, one has to analyze how changing the diagonal of $Q$ influences the result of the eigenproblem.
\section{Methods}

We applied our theoretical framework to two illustrative examples in Sec.~\ref{sec:results}. Here we provide the corresponding algorithmic and implementation details. First, we show how we can use the tSqRA to compute the application of $\Qt$ and thereby also its spectrum without the large memory overhead of the representions of  $Q$ in a dense or sparse format, leading to exponential respectively linear memory savings in the number of dimensions.
Then, we introduce the \emph{Projected Rate Estimation} (PRE), an efficient algorithm to estimate the macroscopical transition rates of a coupled system from comparatively few simulations using the tSqRA of the combined system as a prior.

\subsection{Exploiting the tensor formulation to solve the eigenvalue problem of the coupled system}
\label{sec:tensoreigenproblem}

The use of the tensor formulation does not offer any particular advantage in solving e.g. eigenvalue problem of the combined matrix $Q$, where it is more convenient to compose the the individual eigenvalues, eqs.~\eqref{eq:sum_evals} and eigenvectors \eqref{eq:kron_sum_evecs}. However, in the case of the coupled system $\Qt$ it permits to avoid the explicit construction of a matrix for the application of $\Qt$ to a vector, thus enabling the use of matrix-free methods.

Using the results for a global coupling (Section \ref{sec:theorycoupling}), we write the rate matrix applied to a column vector $v$ of size $(M \times 1)$ as
\begin{align}
    \widetilde{Q} v = D\cind\inv \Qo D\cind v
    - 
    \mathrm{diag}\left[D\cind\inv
    \Qo
    D\cind \, e \right] v \, .
    \label{eq:widetildeQ}
\end{align}
Making use of the definition $D\cind = \mathrm{diag}[\sqrt{\pi\cind}]$ \eqref{box:box_pert} and the fact that multiplying a diagonal matrix from the left is equivalent to the element-wise product $\circ: \mathbb{R}^n \times  \mathbb{R}^n \rightarrow  \mathbb{R}^n$,
\begin{equation}
    (x\circ y)_i :=x_i y_i = (\text{diag}[x] y)_i \, ,
\end{equation}
the left hand term of the matrix-vector product \eqref{eq:widetildeQ} becomes
\begin{align}
    \left[    D\cind\inv \Qo D\cind    \right]  v 
    = \pi\cind^{-\nicefrac{1}{2}} \circ \Qo \left( \pi\cind^{\nicefrac{1}{2}} \circ v\right) \,
    .
 \label{eq:Qv1_1}
\end{align}
%
%
Likewise, for the second term we can write
\begin{align}
    \mathrm{diag}\left[D\cind\inv\Qo D\cind \, e \right] v 
    = \left( \pi\cind^{-\nicefrac{1}{2}} \circ \Qo  \pi\cind^{\nicefrac{1}{2}}\right) \circ v 
  .
 \label{eq:Qv1_2}
\end{align}
In this way we have transformed the application of $\Qt$ \eqref{eq:widetildeQ}
to a series of element-wise operations as well as the application of $\Qo$:
\begin{align}
    \widetilde{Q} v = 
    \pi\cind^{-\nicefrac{1}{2}} \circ \Qo \left( \pi\cind^{\nicefrac{1}{2}} \circ v\right)
    - 
    \left( \pi\cind^{-\nicefrac{1}{2}} \circ \Qo  \pi\cind^{\nicefrac{1}{2}}\right) \circ v.
\end{align}

Note that $\Qo$ is the result of the Kronecker sum of the matrices $Q_i$ and as such can be efficiently stored and computed by applying the individual $Q_i$ to the relevant components of the respective input.

In Appendix \ref{sec:matrixdirectsum} we illustrate how this can be achieved by a series of rearrangements of the input followed by a batched matrix-vector product and provide a corresponding python implementation.

Alternatively to this approach of pulling back the computation to matrix algebra, the banded structure of the Kronecker sum can be used to devise a direct cache efficient implementation of its action. We provide a self-contained Julia implementation of this approach, together with an implementation of the tensor formalism from Section \ref{sec:theorytensor} in Appendix \ref{app:juliacode}

In any case, both approaches allow us to calculate the eigenvalues of $\widetilde{Q}$ using matrix-free solvers like the ARPACK algorithm \cite{lehoucq1997arpack} without ever creating the whole matrix $\widetilde{Q}$.
It follows that we can write the vectors $v$ and $e$ as tensors of order $n$ and shape $(M_1, M_2, \dots, M_n)$, and the two matrix-vector multiplications as sum of tensor dot products as in the previous section .

This has strong implications in terms of memory consumption. Consider a system with $n$ dimensions, discretized into $m$ bins each: A state vector in that case has a size of $8\cdot m^n$ bytes, a dense representation of $\Qt$ would need $8\cdot m^{2n}$ bytes, but even when using a sparse representation it would have approximately $8\cdot 2nm^n$ bytes. For e.g. a system in $n=9$ dimensions with $m=10$ bins each, we would need approximately $8$ GB for the state, $8\times10^{9}$ GB for a dense and $134$ GB for a sparse $\Qt$ matrix.
Using the tensor decomposition of coupled systems the biggest part of storing $\Qt$ is the storage of $D\cind$ which has the same size as a single state vector.

%
%
%
%
%

%

\subsection{The projected rate estimation}
\label{sec:ECTSofIS}

When analyzing a coupled system one is not necessarily interested in the whole matrix $\wt Q$. In practice it may be sufficient to study the impact of the coupling on the characteristic time-scales, such as the rates between metastable macrostates. 

To this end we propose an algorithm that can be understood as hybrid between the formal SqRA and the data-based Koopman estimation by trajectory simulation.
We use the SqRA on the isolated systems to compute their spectrum and obtain the eigenfunctions of the combined systems by means of equation eq.~\eqref{eq:kron_sum_evecs}.
Using the PCCA+ algorithm we transform these into membership functions that characterize the slow-scale dynamics and span an invariant subspace of the combined system.

We then use a Gillespie algorithm to simulate trajectories according to the coupled system (without the need to compute the whole $\widetilde Q$).
By projecting these trajectories onto the membership functions of the combined system we can estimate the slowest rates of the interacting coupled.

To this end, let us quickly introduce PCCA+, an algorithm to compute the characteristic time-scales, a process also known as ``coarse-graining''.
The fundamental object of PCCA+ are the membership functions $\chi \in \mathbb{R}^{M\times n_C}$ obtained by a linear combination of the $n_C$ dominant eigenfunctions $X$ of $Q$ via the matrix $A\in \mathbb{R}^{n_C\times n_C}$,
\begin{equation}
    \chi = XA \,.
\end{equation}
PCCA+ finds a matrix $A$ such that $\chi$ is non-negative and all its row sums equal to one.
The projection of $Q$ onto this subspace leads to the coarse-grained generator $Q_c = A^{-1} \Lambda A
\in \mathbb{R}^{n_C\times n_C}$ that provides the rates between the macroscopic states characterized by $\chi$.
One defining property of $Q_c$ and $\chi$ is that time propagation of the system (via $Q$ or $Q_c$) and projection to the subspace $\chi$ commute, i.e. $Q\chi = \chi Q_c$, or using the (Moore-Penrose) pseudoinverse $Q_c = \chi^+ Q \chi$.
Analogous results hold for the Koopman operator $K = K^\tau = \exp (\tau Q)$ with lag time $\tau\in\mathbb{R}$ (and the coarse grained $K_c$) which has the advantage that it allows for a Monte-Carlo approximation via simulations of the system,

\begin{eqnarray}
    K^{\tau} \chi(x) \approx \frac{1}{N} \sum_{i=1}^{N} \chi(x_i^{\tau}) \, .
    \label{eq:montecarloK}
\end{eqnarray}
where $x_i^\tau$ are the end points of $N$ sampled trajectories of length $\tau$ started in $x$.
The guiding idea of this algorithm is that \begin{equation}
    K \chi = \chi K_c
\end{equation}
and similarly for the coupled system $\wt K \wt \chi = \wt \chi \wt K_c$ if $\wt \chi$ was computed with respect to $\wt K$, which in practice is not known.

However, for small coupling terms, we can expect that the invariant subspace of the combined system would not change too much, i.e. $\chi \approx \wt \chi$. This then leads to 

\begin{equation}
\label{eq:approxQc}
    \wt K \chi \approx \chi \wt K_c
\end{equation}
which then allows us to approximate the coarse grained coupled dynamics via 
\begin{equation}
\label{eq:linsys}
    \hat K_c = \chi^+ \wt K \chi \approx \wt K_c.
\end{equation}
and the Monte-Carlo approximation \eqref{eq:montecarloK} of $\wt K \chi$ using the coupled dynamics.

Given that $n_C < M$, the problem for solving for $\hat K_C$ is inherently overdetermined.
We can take advantage of this by estimating only a few rows of $K \chi$ (and the corresponding columns of $\chi^+$).
This goes in hand with a greatly reduced amount of required simulations and is the source of the efficiency of this method.

Taking the matrix logarithm then allows us to recover the approximated rates of the coupled system $\hat Q_c = \mathrm{logm}\, \hat K_c$. We obtained better results by re-normalizing the rows of $\hat K_c$ to have row sum one, corresponding to the interpretation of a probability matrix. We expect that the results could be even further improved by directly restricting the solution of the linear system \eqref{eq:linsys} to the unit simplex using constrained optimization (with optimization objective \eqref{eqn:linsysopt}).

The algorithm to estimate $\Qt_c$ is summarized as follows:
\begin{itemize}
\item[1.] Solve the eigenproblems for the isolated sub-systems:
\begin{eqnarray}
    Q_i X_{i} = \lambda_{i} X_{i} \, ,
    \label{eq:eigenvalueProblems}
\end{eqnarray}
where $X_{i}$ are the matrices containing the eigenvectors of the $i$th subsystem, and $\lambda_i$ are the vectors containing the corresponding eigenvalues.
\item[2.] Identify the number $c_i$ of metastabilities for each $i$th subsystem from the analysis of the eigenspectrum and compute the membership functions $\chi_i$ via PCCA+.
\item [4.] Compute the membership functions of the combined system
\begin{eqnarray}
    \chi = \bigotimes\limits_{i=1}^n \chi_i \, .
\end{eqnarray}
Note that $\chi_i$ is a matrix of size $M_i \times c_i$, while $\chi$ is a matrix of size $\prod_i M_i \times \prod_i c_i$.
\item[5.] Choose $n$ starting points $x=(x_1,x_2,...,x_n)$, $n_c<n\ll M$ for the Gillespie simulations.
\item[6.] From each of these points $x$, start $N_{R}$ replicas of Gillespie simulations of length $\tau$ using the transition rates of the coupled SqRA, $\wt Q$, which needs to be computed only locally, i.e. row-wise.
\item[7.] Compute the (partial) $\wt K \chi \in \mathbb{R}^{n \times M}$
\begin{eqnarray}
    \left[ \widetilde{K} \chi \right] _{ij} 
    \approx \frac{1}{N_R} \sum_{n=1}^{N_{R}} \chi_j(x_{i}^{n}) \, .
\end{eqnarray}
where $x_{i}^{n}$ is the end point of the $n$-th Gillespie simulation starting at $x_i$.
%
\item[8] Compute the pseudoinverse to estimate $\hat K_c = \chi^+\wt K \chi$
\item[9.] Normalize the rows of $\hat{K}_c$.
\item [10.] Estimate the approximated coarse-grained rate matrix as
\begin{eqnarray}
    \hat{Q}_c = \frac{1}{\tau} \mathrm{logm}\,  \hat{K}_c \, ,
\end{eqnarray}
where $\mathrm{logm}(\cdot)$ denotes the matrix logarithm.
\end{itemize}

\section{Results}\label{sec:results}
\subsection{Two-Dimensional Example}

This 2-dimensional example illustrates our theoretical elaborations. No special increases in efficiency are required to recalculate these examples in a comprehensible manner.
\subsubsection{Numerical experiment parameters}
As an illustrative example, we considered the overdamped Langevin dynamics of a particle of mass $m = 1\, \mathrm{amu}$ and friction $\xi = 1\, \mathrm{ps}^{-1}$ which moves in a two-dimensional space under the action of the potential energy function
\begin{eqnarray}
    \widetilde{V}(x_1, x_2) 
    &=& 
    V_1(x_1) + V_2(x_2) + c V_{12}(x_1, x_2) \cr 
    &=&
    (x_1^2 - 1)^2 + x_1 + 2 x_2^2 + c\,x_1 x_2 \quad [\mathrm{kJ\,mol^{-1}}] \, .
    \label{eq:Potential2D}
\end{eqnarray}
The function is made of two potentials $V_1$ and $V_2$ describing the dynamics of two non-interacting subsystems along the coordinates $x_1$ and $x_2$ respectively.
Additionally, a coupling term $V_{12}$ can be activated ($c\neq 0$) or deactivated (c=0) by the parameter $c$.
The two-dimensional function, illustrated in fig.~\ref{fig:fig1}-(a) for $c=0$ and $c=1$, describes a surface with two wells of different heights separated by a barrier.

For our numerical experiments, we assumed standard thermodynamic parameters:
the temperature of the system was $T=300\, \mathrm{K}$, the molar Boltzmann constant $k_B = 0.008314463 \,\mathrm{kJ\cdot mol^{-1}\cdot K^{-1}}$, and the diffusion constant was $D = 2.49\,\mathrm{nm^2\,ps^{-1}}$ in each direction.

\subsubsection{The rate matrix}
In order to build the rate matrix by SqRA, we discretized the $x_1$-range $[-3.4,\,3.4]$ nm and $x_2$-range $[-3.4,\,3.4]$ nm respectively, in $M_1 = M_2 = 50$ subsets of the same length $\Delta x_1 = \Delta x_2 = 0.13$ nm, for a total of $M = M_1\times M_2 = 2500$ square subsets of the two-dimensional space.

For the combined system ($c=0$), we built respectively the rate matrices $Q_1$ and $Q_2$ (size $50\times 50$), then, we estimated the first four eigenvalues and right eigenvectors solving the eigenvalue problems
\begin{eqnarray}
    Q_1 X_{1} = \lambda_{1} X_{1} \, ,
\end{eqnarray}
and
\begin{eqnarray}
    Q_2 X_{2} = \lambda_{2} X_{2} \, ,
\end{eqnarray}
where $X_{1}$ and $X_2$ are matrices of size $50\times 4$ containing the first four eigenvectors, and $\lambda_1$ and $\lambda_2$ are four-dimensional vectors containing the corresponding eigenvalues.
For the combined system ($c=0$), the rate matrix $Q$ of the entire system can be estimated from the Kronecker sum of the rate matrices $Q_1$ and $Q_2$.
Exploiting this property (see Ref.~\cite{Loan2000} for more details), we estimated sixteen eigenvalues $\lambda$ of $Q$ as the sum of the eigenvalues of $Q_1$ and $Q_2$ (eq.~\eqref{eq:sum_evals}),
\begin{eqnarray}
    \lambda_k = \lambda_{1,i} + \lambda_{2,j} \quad \forall i,j = 0,1,2,3 \, ,
    \label{eq:lambda_k}
\end{eqnarray}
and the corresponding eigenvectors $X$ as Kronecker product (eq.~\eqref{eq:kron_sum_evecs}),
\begin{eqnarray}
    X_k = X_{1,i} \kron X_{2,j} \quad \forall i,j = 0,1,2,3 \, .
    \label{eq:X_k}
\end{eqnarray}
In eqs.~\eqref{eq:lambda_k} and \eqref{eq:X_k}, we introduce the index $k=i+4j$.
Note that the eigenvectors $X_1$ and $X_2$ have size $M_1 = M_2 = 50$, while $X$ has size $M = 2500$.

\begin{figure*}[ht!]
    \centering
    \includegraphics{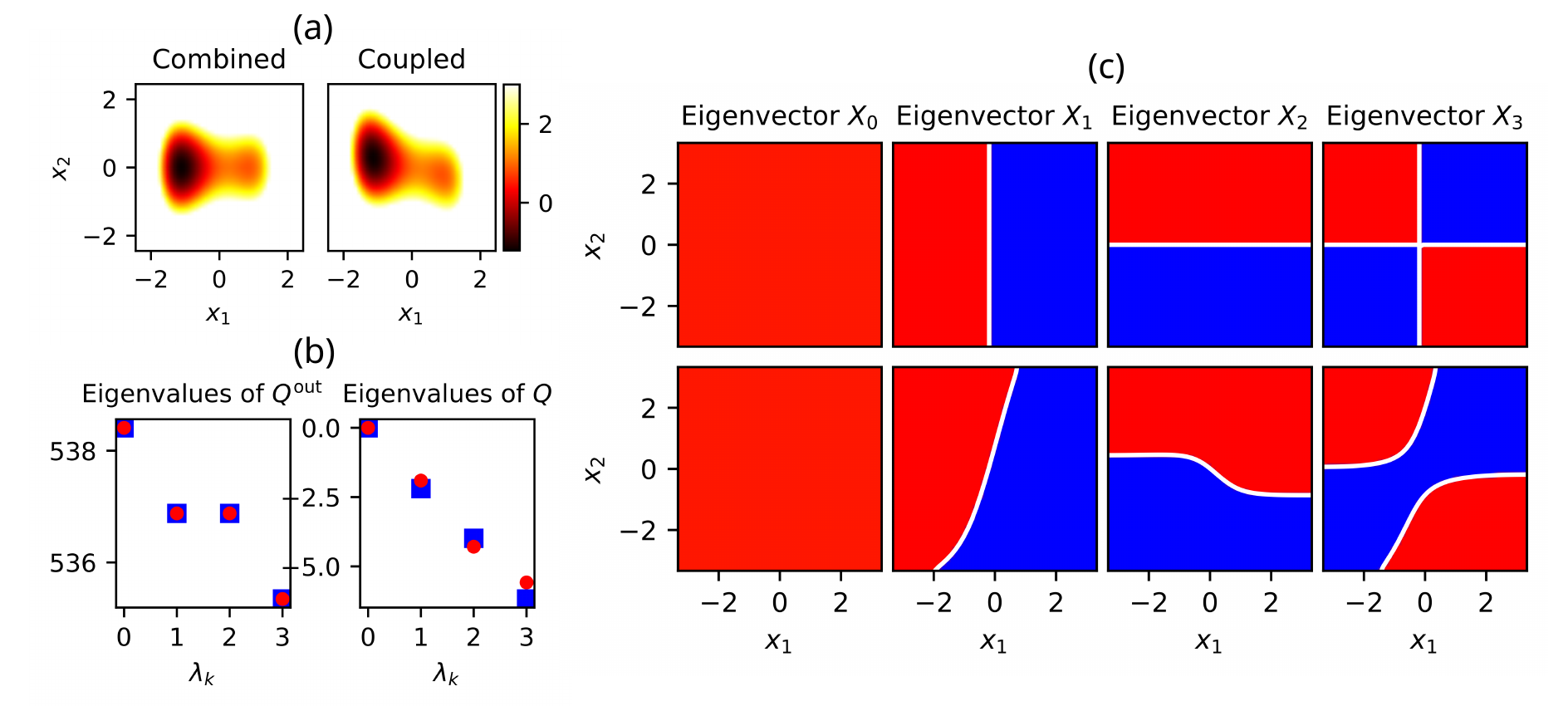}
    \caption{2D model. 
    (a) Potential energy function for combined (left) and coupled system with $c=1$ (right); 
    (b) Eigenvalues of the matrix $\Qo$ (left) and $Q$ (right) for the combined (blue) and coupled system (red);
    (c) First four eigenvectors for combined (top) and coupled system (bottom), the red and blue color denotes positive and negative values respectively, the white lines represent the zero-values.}
    \label{fig:fig1}
\end{figure*}

The eigenvalues and the corresponding eigenvectors are reported respectively in fig.~\ref{fig:fig1}-(b,c).
The first eigenvector is constant and is associated to the eigenvalue $\lambda = 0$.
This eigenvector represents the stationary state of the system when the other eigenmodes have decayed.
The other eigenvectors $X_k$, with $k>0$, represent the dominant kinetic processes between regions of the space with negative (blue color) and positive values (red color).
The associated eigenvalues $\lambda_k$ are negative and represent the timescales $-\nicefrac{1}{\lambda_k}$ at which the dominant processes decay.
We conclude that the slowest process is a transition between the region of the space with $x_1<0$ and $x_1>0$, the second slowest process is a transition between the region of the space with $x_2<0$ and $x_2>0$ and the third slowest process is a mix transition between the quadrants of the plane.

For the coupled system ($c\neq 1$), the relations in eqs.~\eqref{eq:lambda_k} and \eqref{eq:X_k} do not hold, i.e. the eigenvectors and eigenvalues of the full system are not directly derived from  those of the individual subsystems.
In this example, we built the full matrix $\widetilde Q$ applying eq.~\eqref{eq:Qcoup} which requires to provide the diagonal matrix $D_{12}$ of size $50\times 50 = 2500$.
The main effect of the coupling term is to distort the eigenvectors and to break the symmetries of the combined system.
With regard to the eigenvalues, we observe a rise in the second eigenvalue $\lambda_1$ and a decrease in the third eigenvalue $\lambda_2$.
From a physical perspective, this corresponds to an acceleration of the transition along the $x_2$ coordinate and a slowdown along the $x_1$ coordinate.
In fig.~\ref{fig:fig1}-(b), we also plot the eigenvalues of the off-diagonal matrices $\Qo$ and $D_{12}\inv\Qo  D_{12}$.
The eigenvalues of the two matrices perfectly match, due to a similarity transform. 
Thus, it is the matrix $\widetilde E$ of the total leaving rates defined in eq.~\eqref{box:box_pert} that affects the timescales of the system.

\subsubsection{Coarse-grained rate matrix}
The analysis of the system's eigenvectors and eigenvalues indicates that the system oscillates between two metastable states, i.e. two regions of the state space where the system resides for most of the time, with the transition from one state to the other occurring rarely.
It is therefore convenient to reduce the matrices $Q$ and $\widetilde{Q}$ to matrices $Q_c$ and $\widetilde{Q}_c$ of size $2\times2$, containing the rates $k_{12}$ and $k_{21}$ between the two metastable regions.
For this purpose, we used PCCA+, which provides the membership functions $\chi$ containing the probabilities that a state belongs to the identified metastable states.
The membership functions can then be used to build the coarse-grained matrix $Q_c$ by projecting the fine-grained matrix $Q$:
\begin{eqnarray}
Q_c
&=&
(\chi^\top \mathrm{diag}(\pi) \chi)^{-1} \chi^\top \mathrm{diag}(\pi) Q \chi  \, .
\label{eq:Qc}
\end{eqnarray}
With regard to the combined system ($c=0$), in practice, it is more convenient to exploit the algebraic structure of the rate matrix $Q$ and to apply PCCA+ to $Q_1$ and $Q_2$ to obtain the membership functions $\chi_1$ (size $50\times 2$) and $\chi_2$ (size $50\times 1$), 
and the rate matrices $Q_{c,1}$ (size $2\times 2$) and $Q_{c,2}$ (size $1$).
Note that $\chi_2 = \mathbf{1}$ (size 50) because the system has one metastability along the $x_2$ direction and $Q_{c,2} = 0$.
Then, the membership functions of the coupled system are obtained by the Kronecker product
\begin{eqnarray}
    \chi = \chi_1 \kron \chi_2 \, ,
\end{eqnarray}
while the rate matrix is given by the Kronecker sum
\begin{eqnarray}
    Q_c = Q_{c,1} \oplus Q_{c,2} = Q_{c,1} =
    \begin{pmatrix}
    -1.45 & 1.45 \cr
    0.74  & -0.74
    \end{pmatrix} \, .
\end{eqnarray}
Because the system has two metastabilities, the Kronecker product simply results in an increase in the size of $\chi_1$ from 50 entries to $50 \times 50=2500$ entries since $\chi_2$ is a constant function, and the rate matrix $Q_c$ is equal to $Q_{c,1}$.
However, if the second subsystem has two metastabilities as well, then the coupled system would have four metastabilities and the corresponding $Q_c$ matrix would have the size $4\times 4$.

\subsubsection{Coupled system}
\begin{figure*}[ht!]
    \centering
    \includegraphics{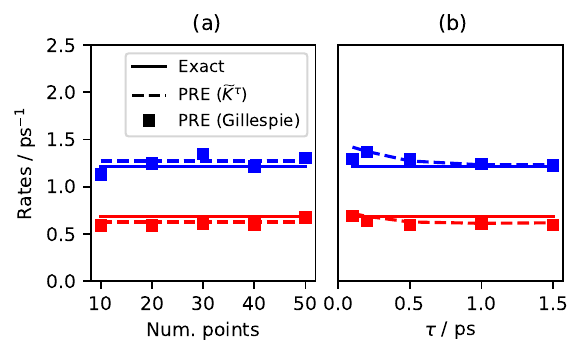}
    \caption{2D model. Transition rates $k_{12}$ (blue) and $k_{21}$ (red) of the coarse-grained model as functions of the number of starting points (a) and the lag time $\tau$ (b).
    The solid lines represent the exact results using $\Qt$, the dashed lines represent the projected rate estimation (PRE) using the exact $\widetilde{K}^{\tau} = \exp (Q\tau)$, and the squares represent the PRE using Gillespie's algorithm to estimate $\widetilde{K}^{\tau}$ .
    }
    \label{fig:fig2}
\end{figure*}

For the coupled system ($c=1$) we applied the PRE algorithm descrived in section \ref{sec:ECTSofIS}.
The algorithm is based on the assumption that the Koopman operator $\widetilde{K}^{\tau}$ of the coupled system applied to membership functions $\chi$ of the combined system, well approximates the resultant of the coarse-grained Koopman operator $\widetilde{K}^{\tau}_c$ applied to $\chi$ (see eq.~\eqref{eq:approxQc}).
First of all, we verified the validity of this assumption by constructing the discretized coupled Koopman operator as
\begin{eqnarray}
    \widetilde{K}^{\tau} = \mathrm{expm}\left(\tau \widetilde{Q} \right) \, ,
\end{eqnarray}
where $\widetilde Q$ is the rate matrix of the coupled system built by SqRA and $\mathrm{expm(\cdot)}$ denotes a matrix exponential.
The matrix $\widetilde{K}^{\tau}$ has been multiplied to the membership functions $\chi$ of the combined system and we solved the linear regression problem
\begin{eqnarray}
\label{eqn:linsysopt}
    \min_{\hat{K}_c} \Vert \widetilde{K}^{\tau} \chi - 
    \chi \hat{K}^{\tau}_c \Vert \, ,
    \label{eq:linReg}
\end{eqnarray}
i.e. we looked for the matrix $\hat{K}^{\tau}_c$ (size $2\times 2$) that minimizes the 2-norm $\Vert \widetilde{K}^{\tau} \chi -  \chi \hat{K}^{\tau}_c \Vert$.
The matrix $\hat{K}^{\tau}_c$ approximates the exact coarse-grained Koopman operator $\widetilde{K}^{\tau}_c$ for the coupled system and yields the approximated coarse-grained rate matrix as
\begin{eqnarray}
    \hat{Q}_c = \frac{1}{\tau} \mathrm{logm}\,  \hat{K}^{\tau}_c \, .
\end{eqnarray}
In fig.~\ref{fig:fig2}-(b), we plotted the off-diagonal entries of $\hat{Q}_c$ (dashed lines) for several $\tau$ values in the range [0,1.5] ps, and compared with the exact values of $\widetilde{Q}_c$ (solid lines) obtained by applying PCCA+ to $\widetilde{Q}$.
We observe that while the exact results do not depend on the choice of lag time $\tau$, the approximate results tend to converge only for large values of $\tau$.
However, the approximate rates do not reach the exact values.

If building the matrix $\widetilde{K}^{\tau}$ is not feasible due to the high dimensionality, the action of the Koopman operator on the membership functions $\chi$ can only be approximated for some points $x$ in the state space.
Provided $N_{R}$ trajectories of length $\tau$, starting in $x$ and reaching the points $x_i^{\tau}$, with $i=1,2,...,N_{R}$ yields
\begin{eqnarray}
    \widetilde{K}^{\tau} \chi(x) \approx \frac{1}{N_{R}} \sum_i^{N_{R}} \chi(x_i^{\tau}) \, .
    \label{eq:appKoop}
\end{eqnarray}
The trajectories could be generated by solving the underlying equations of motion of the coupled system, for example by using the Euler-Maruyama integration scheme.
Instead, we exploited the knowledge of the transition rates between adjacent subsets of the state space, approximated by SqRA, and applied the Gillespie algorithm.

In order to explain the procedure, consider the following example.
Let us assume that we know the matrices $Q_1$ and $Q_2$ of the one-dimensional individual subsystems and that the system is in a state $x=(x_1,x_2)$ belonging to a subset of the state space with indices $(i_1,i_2)$.
The system, in an infinitesimal interval of time, can evolve its state in five ways: two transitions along the $x_1$ direction, two along the $x_2$ direction, or no transition.
The four transition rates are contained in the $Q_1$ and $Q_2$ matrices, but these must be reweighted according to the coupling term $V_{12}$ in eq.~\eqref{eq:Potential2D}.
For example, the forward transition rate along the $x_1$ direction becomes:
\begin{eqnarray}
    \widetilde
    {Q}_1^{i_1,i_1+1} 
    = 
    Q_1^{i_1,i_1+1} 
    \cdot
    \frac
    {\sqrt{\pi_{12}^{i_1+1,i_2}}}
    {\sqrt{\pi_{12}^{i_1,i_2}}} \, .
\end{eqnarray}
Likewise, we obtain the rates $\widetilde{Q}_1^{i_1,i_1-1}$, $\widetilde{Q}_2^{i_2,i_2+1}$ and $\widetilde{Q}_2^{i_2,i_2-1}$ and the total leaving rate given by minus the sum of the former.
Thus, we have estimated only a row of the matrix $\widetilde{Q}$, which can be used within a standard Gillespie algorithm to estimate: (i) the infinitesimal time interval at which the next transition occurs, (ii) which of the five possible events (four transitions and no transition) takes place.
The reweighting of rates for the coupled system occurs at each simulation step, however, for highly dimensional systems this is the only solution since it is not possible to estimate the entire matrix $\widetilde{Q}$.

Since the Gillespie algorithm provides the solution in terms of jumps between the subsets of the state space, it provides the indices $(j_1,j_2)$ of the arrival subset of $x^{\tau}$ and the calculation of the membership functions can be approximated by calculating the value that $\chi$ assumes in the target subsets:
\begin{eqnarray}
    \chi(x^{\tau}) \approx \chi^{j_1,j_2} \, .
\end{eqnarray}
Repeating this calculation for $N_{R}$ trajectories, one can approximate the Koopman operator according to eq.~\eqref{eq:appKoop} and solve the linear regression problem in eq.~\eqref{eq:linReg}.
However, unlike the situation where we have the entire matrix $\widetilde{K}^{\tau}$, after solving the linear regression problem, it is necessary to normalize the rows of the matrix $\widetilde{K}^{\tau}_c$.

We studied the sensitivity of the method to the initial points and the lag time $\tau$ of the Koopman operator and reported the results in fig.~\ref{fig:fig2}.
First of all, we studied the dependence of the method on the number of initial points $x$, randomly drawn from a uniform distribution that covers the state space, setting a lag time equal to $\tau=0.5$ ps and a number of replicas per point $N_{R}=100$.
We do not observe any particular improvement in the results by increasing the number of initial points, so for the next analysis, we considered 10 initial points and increased the number of replicas to 500 to reduce the stochastic error.
Excluding the first point, which corresponds to a lag time of 0.1 ps, the agreement with the expected results (dashed line) is excellent.
%

%
%

\subsection{$n$-dimensional example}

Without special mathematical tricks one could not calculate the SqRA for high-dimensional examples. In this example, we heavily use the algebraic structure of tSqRA to rigorously calculate the eigenvalues of a large rate ``matrix''. That this becomes possible is shown in Fig.~\ref{fig:fig3}.

\subsubsection{Numerical experiment parameters.}
\begin{figure*}[ht]
    \centering
    \includegraphics{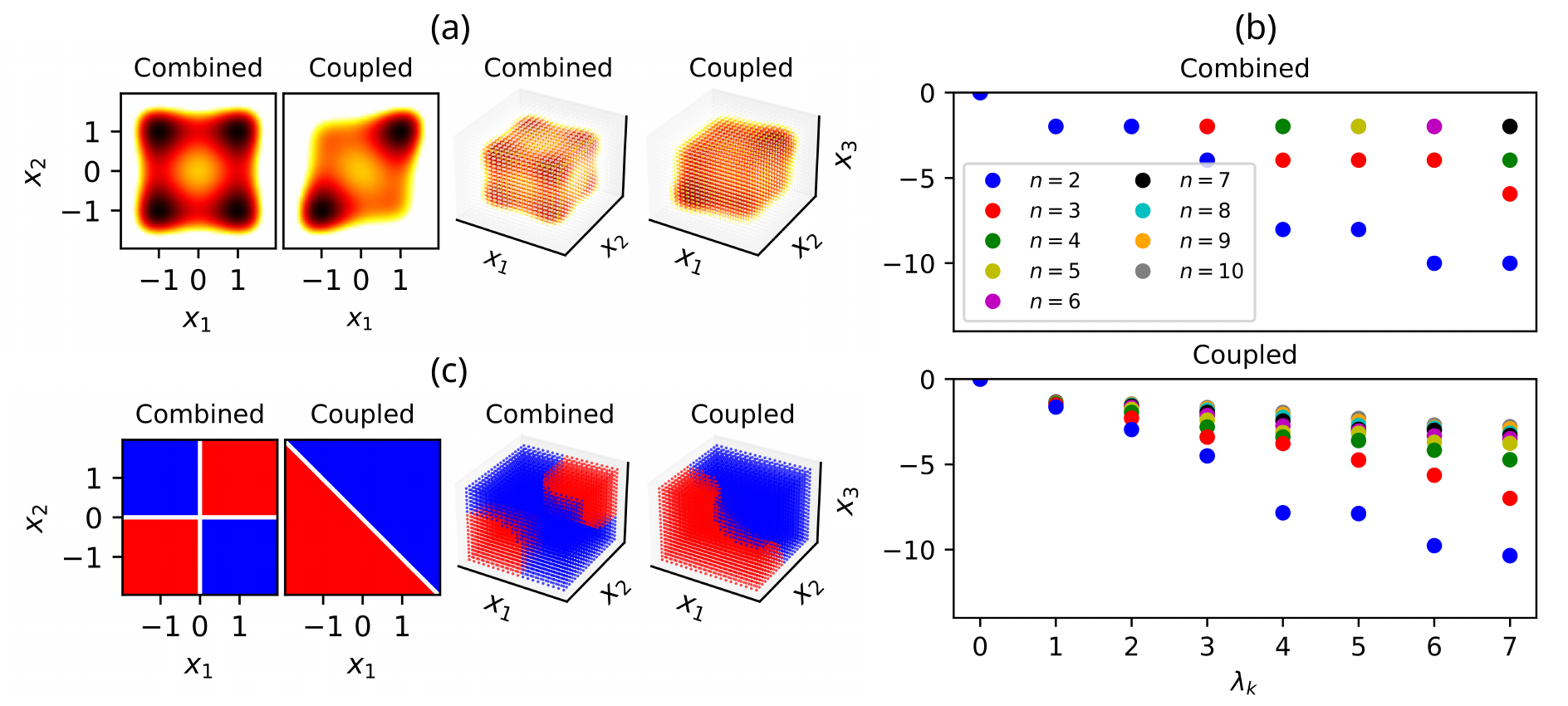}
    \caption{
    Metastable chain. 
    (a) Combined and coupled potential energy function for $n=2$ and $n=3$, dark and bright colors represent respectively low and high energy values.;
    (b) first 8 eigenvalues for $n=2,3,\dots 10$ of the combined and coupled system;
    (c) second eigenvectors for $n=2$ and $n=3$ of the combined and coupled system.
    }
    \label{fig:fig3}
\end{figure*}

As a second example, we studied a ``metastable chain'' made of $n$ bimetastable systems interacting via harmonic oscillators.
The potential energy function describing the dynamics of this system is written as
\begin{eqnarray}
\widetilde{V}(x_1,x_2,\dots,x_n)
&=& 
V_1(x_1) + V_2(x_2) + \dots + V_n(x_n)
\cr
&&+c\cdot \left[
V_{12}(x_1,x_2) + V_{23}(x_2,x_3) + \dots + V_{n-1,n}(x_{n-1},x_n)
\right]\cr
&=& 
\sum_{i=1}^{n} V_i(x_i)
+
c\cdot
\sum_{i=1}^{n-1} V_{i,i+1}(x_i,x_{i+1}) 
\, ,
\end{eqnarray}
where 
\begin{eqnarray}
    V_i(x_i) = (x_i^2 - 1)^2 \, ,
\end{eqnarray}
and
\begin{eqnarray}
    V_{i,i+1}(x_i,x_{i+1}) = \frac{1}{2}\left\vert x_i - x_{i+1} \right\vert^2 \, .
\end{eqnarray}
The combined system ($c=0$) is characterized by $n^2$ metastable states, while the coupled system ($c=1$) has only two metastable states, as illustrated in fig.~\ref{fig:fig3}-(a) for $n=2$ and $n=3$, respectively.
For the numerical experiments, we used the same thermodynamic parameters as in the previous examples.
To build the rate matrix $Q_i$ of each bimetastable system by SqRA, we discretized the range $[-2.5,\,2.5]$ nm of each coordinate $x_i$ into $M_i = 5$ subsets of length $\Delta x_i = 1.0$ nm, then the total number of subsets of the system is $5^n$.

Differently from the previous example, constructing the rate matrices $Q$ and $\widetilde{Q}$ of the coupled system (combined and coupled) is not feasible for high values of $n$.
Thus, we used the approach described in section \ref{sec:tensoreigenproblem} to calculate the eigenvalues and eigenvectors without building the full rate matrices but providing to the eigensolver a function that represents their action on a vector of size $(M \times 1)$.
Fig.~\ref{fig:fig3}-(b) shows the first 8 eigenvalues of the rate matrix $Q$ and $\widetilde{Q}$.
For the combined case (a), we observe that the number of eigenvalues associated to the slowest processes is equal to $n$, then the slowest transitions occur along the coordinates $x_i$.
On the contrary, the effect of the coupling term is to change the symmetry of the system and to create a hierarchy of sub-processes arising at different time scales.
Thus, we observe a decreasing scale of eigenvalues for each $n$-dimensional system.
The first non-zero eigenvalue corresponds to the slowest process occurring along the diagonal of the $n$-dimensional circle, as shown in the fig.~\ref{fig:fig3}-(c) by the corresponding eigenvectors for the case with $n=2$ and $n=3$.
%

%
%
%
%
\section{Discussion and conclusion}
\label{sec:disc_sec}
%

This article shows how the algebraic structure of SqRA can be exploited in order to systematically analyze the macroscopic timescales of slow processes in coupled molecular systems from first principles. Our article is a step towards understanding the origins of these timescales and the influence of interaction energies between subsystems.

Using the tensor approach (tSqRA) we can describe the direct relation between energies and rates of coupled systems which allows for its mathematical analysis based on first principle models of molecular processes.
Combining the tSqRA with PCCA+ we also suggest a new hybrid method, the projected rate estimation (PRE), which uses the combined non-interacting subsystems (model-based) as prior for an effective estimation of the macroscopical rates of a coupled system by a comparatively small number of simulations (data-based).


On the computational side, the tSqRA improves the applicability and efficacy of the SqRA for coupled systems even without the use of low-rank approximations.
However, we also explain how tSqRA can be extended to incorporate low-rank approximations.
Currently the limiting factor for an efficient tensor-based calculation is the dense representation of the system's state, which is succumbed to the curse of dimensionality.
However, a low rank representation, e.g. in the form of tensor networks, could alleviate that problem \cite{Cichocki_2016}.
The lacking piece so far is the support for truncated Hadamard-products. 

Note in this regard that the adjacency matrix $A$, being a Kronecker sum, is inherently low rank. Similarly the matrix $D$ is composed of Hadamard products like in eq.~\eqref{eqn:dproduct}.
Assuming that the system under consideration admits a low-rank representation, as to be expected for two weakly coupled systems, we expect the exploitation of these low-rank structures to dramatically improve both memory and compute costs.


From our perspective, it is worthwhile to better comprehend the computational advantage of tensor-based complexity reduction methods for future analysis of molecular systems without having to generate molecular simulation data. 

\section*{Data availability statement}
The scripts for recreating the experiments as well as from the appendix can be found at \href{https://github.com/zib-cmd/article-tsqra}{https://github.com/zib-cmd/article-tsqra}.

\section*{Acknowledgement} \theacknowledge 

\printbibliography

\newpage
\appendix
\section*{Appendix}

\section{Python implementation of a global coupling term}
\label{sec:appendix1}

\subsection{From matrix-vector multiplication to sum of tensor dot products
}
\label{sec:matrixdirectsum}
When computing with tensors one difficulty is how to actually represent them for computations, either as $n$-dimensional arrays or in their "flattened" representation, where states are vectors and operators matrices.
In this section we describe how we can compute the action of a direct sum of matrices, interpreted as flattened matrix, as the sum of matrix products along the individual tensor modes.  
We will then use this approach in the following Python implementation.

Consider $n$ matrices $Q_i$ each of same size $(M_i \times M_i)$ and the matrix $Q$ of size $(M_i^n \times M_i^n)$ obtained as Kronecker sum of $Q_i$ as in eq.~\eqref{eq:Qdeco}. 
Given a arbitrary column vector $v$ of size $(M_i^n \times 1)$, the result of the matrix-vector multiplication is the column vector
\begin{equation}
    u = Q v = \left( \bigoplus_{i=1}^n Q_i \right) v     \, ,
    \label{eq:matvecmult1}
\end{equation}
of size $(M_i^n \times 1)$.
This operation requires the explicit construction of the matrix $Q$.
Instead, we now reformulate the same operation as a sum of tensor dot products applied to the smaller matrices $Q_i$.

First, we reshape the vector $v$ into a tensor of order $n$ with shape $(M_1,M_2,\dots,M_n)$ and entries $v^{\gamma_1,\gamma_2,\dots,\gamma_n}$, where each index $\gamma_i$ specifies the $i$th dimension of the tensor.
Next, we perform the following two operations on $v$ to build $n$ tensors $u_i$ of order $n$:
\begin{enumerate}
    \item Tensor contraction between the matrix $Q_i$ and the tensor $v$:
    \begin{eqnarray}
    u_i^{
    \alpha, \gamma_1, \gamma_2, 
    \dots,
    \gamma_{n}
    } 
    = 
    \sum_{\kappa_i  = 1}^{M_i}  
    Q_{i}^{\alpha \kappa_i} 
    \, 
    v^{
    \gamma_1, \gamma_2, 
    \dots, \kappa_i, \dots, 
    \gamma_{n}
    }
    \, ,
    \label{eq:TensorContraction}
    \end{eqnarray}
    where $\kappa_i$ is the contraction index, that is, the index on which to sum the products between the entries of the matrix $Q_{i}$ and the entries of the tensor $v$.
    In other words, the entries of the resulting tensor $u_i$ are the dot products between the $k_i$th column of $Q_{i}$ and the $k_i$th dimension of $v$.
    Here, we added the subscript $i$ to $\kappa_i$ to remark that it replaces the $i$th index of $v$ in eq.~\eqref{eq:TensorContraction}.
    It follows that $u_i$ is a tensor of order $n$ which contains all the indexes of $Q_{i}$ and $v$ but the second index of $Q_{i}$ and the $i$th index of $v$.
    The operation in eq.~\eqref{eq:TensorContraction} is repeated $\forall i=1,2,\dots,n$. 
    %
    %
    \item Index swapping:
    \begin{eqnarray}
    u_i^{
    \alpha, \gamma_1, \gamma_2, 
    \dots,
    \gamma_{n}
    }
    \rightarrow
    u_i^{
    \gamma_1, \gamma_2, 
    \dots,\alpha_{(i)},\dots
    \gamma_{n}
    } \, ,
    \end{eqnarray}
    where the notation $\alpha_{(i)}$ remarks that the tensor must be reordered such that $\alpha$ occupies the $i$th position. 
\end{enumerate}
Once we computed the $n$ tensors $u_i$, one for each matrix $Q_i$, the tensor $u$ representing the matrix multiplication between $Q$ and $v$ is the sum of tensors $u_i$ with entries
\begin{eqnarray}
u^{
\gamma_1, \gamma_2,
\dots,
\gamma_n
}
= \sum_{i=1}^n 
u_i^{
    \gamma_1, \gamma_2, 
    \dots,\alpha_{(i)},\dots
    \gamma_{n}
    } \, .
\end{eqnarray}
The final result is a tensor of order $n$ which can be flattened into a one-dimensional vector equal to the result of the matrix-vector multiplication defined in eq.~\eqref{eq:matvecmult1}.
We implement this approach in the following python code.

\subsection{Python codes for computations of $Qv$ and eigenvalues of $\widetilde{Q}$ based on matrices}\label{app:P}
\definecolor{keywords}{rgb}{0,0,0.5}
\definecolor{comments}{RGB}{0,110,113}
\definecolor{red}{RGB}{160,0,0}
\definecolor{green}{RGB}{0,150,0}

\lstset{language=Python, 
        basicstyle=\small\ttfamily\linespread{0.8}\selectfont,
        keywordstyle=\color{green},
        commentstyle=\color{comments},
        stringstyle=\color{red},
        showstringspaces=false,
        identifierstyle=\color{black},
        procnamekeys={def,class}} 
\begin{lstlisting}[language=Python,
caption = {
The code defines the function $\textsf{sum\_tensor\_dots()}$ that calculates the matrix-vector multiplication $Qv$ as sum of tensor dot products, then it solves the eigenvalue problem.
In this example, the matrix $Q_i$ and the vector $v$ are randomly generated, and it is assumed that each dimension is discretized into the same number of subsets.
The function $\textsf{np.tensordot()}$ applies the tensor contraction to axis 1 of the matrix, and the index $i$ of the vector in its tensor formulation. Similar solutions can be also found with the function $\textsf{np.einsum()}$.
Note that in Python the indexing starts from 0.},label={lst1}]

import numpy as np
import scipy.sparse.linalg as spl
n     = 5
Mi    = 2
M     = Mi ** n
Qi    = np.random.normal( 0, 1, ( Mi, Mi ) ) 
v     = np.random.normal( 0, 1, ( M,  1  ) )

tensor_shape = [Mi for i in range(n)]

def Q_matmul(v):
    v            = v.reshape(tensor_shape)
    ui           = np.empty(n, dtype=object)
    for i in range(n):
        td   = np.tensordot( Qi, v, axes = [[1], [i]] )
        for j in range(i):
            td = td.swapaxes(j,j+1)
        ui[i] = td
    return np.sum(ui).flatten('C')
    
u = Q_matmul(v)
lin_op_Q     = spl.LinearOperator((M, M), matvec=Q_matmul)
evals, evecs = spl.eigs(lin_op_Q, k=4, which='LR')
\end{lstlisting}

\begin{lstlisting}[language=Python,
caption = {
The code calculates the eigenvalues of the coupled matrix $\widetilde{Q}$ exploiting the tensor formulation of its action on a vector $v$. The array $\textsf{p}$, of shape $(M_1, M_2, \dots, M_n)$, contains the entries of $\pi_{12\dots n}^{\nicefrac{1}{2}}$.
},label={lst2}]

def tildeQ_matmul(v):
    e     = np.ones(tensor_shape)
    term1 = p ** -1 * Q_matmul( p * v )
    term2 = Q_matmul( p * e ) * v
    out   = term1 - term2
    return out.flatten('C')

lin_op       = spl.LinearOperator((M, M), matvec = tildeQ_matmul)
evals, evecs = spl.eigs(lin_op, k=11, which='LR')

\end{lstlisting}

\section{Julia implementation for multiple low-order interactions}

\label{app:juliacode}
\begin{figure*}
    \centering
    \includegraphics{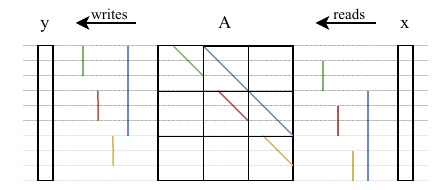}
    \caption{
    Schematic illustration of the cache efficient application of $y=Ax$ exploiting the banded structure for linear memory accesses without storage of A. For simplicity we only show the computations for the upper diagonal, the lower follows analogously.}
    \label{fig:fig4}
\end{figure*}

The following code illustrates how to compute the application of $Q$ using the tensor approach developed in section \ref{sec:theorytensor}.
\verb|apply_A| exploits the fact that A is banded (with entries $1$) to copy and add the specific bands of the input $x$ to the output $y$ (see Figure~\ref{fig:fig4}).
\verb|compute_D| then loops over a given list of potential functions, evaluates them at the grids of the specified dimensions and multiplies them according to eq.~\eqref{eqn:dproduct} along the corresponding modes using Julia's broadcasting capabilities.
We finally compute D and E for the example system with potential $V(x,y) = x^2 + (xy)^2$ on a a square grid of size $10 \times 10$.
The last two functions implement the application of $A-E$ and $Q$ respectively.

Using this we were able to compute a simplified 9-dimensional pentane molecule with 10 cells in each dimension, resulting in a memory demand of $10^9\times 64 bit = 8 Gb$ per state. Whilst the computation of $D$ is a matter of minutes, the computation of the spectrum using an Arnoldi method requires computation time on the order of $1$ day due to the dense state representation.

\begin{lstlisting}[language=Julia,    
    % basicstyle       = \ttfamily,
    keywordstyle     = \bfseries\color{blue},
    stringstyle      = \color{magenta},
    commentstyle     = \color{ForestGreen},
    caption = {Julia implementation of the tSqRA for multiple low-order interactions as in Section \ref{sec:theorytensor}}
    ]
Vec = Vector
Tensor = Array
Grid = Union{Vector,AbstractRange}

function apply_A(x::Tensor, dims::Tuple)
    y = zeros(size(x))
    len = length(x)
    off = 1 # offset
    for cd in dims # current dimension length
        bs = off * cd # blocksize
        for i in 1:bs:len-off # blockstart
            to = i:i+bs-off-1
            y[to] .+= x[to.+off]
            y[to.+off] .+= x[to]
        end
        off = bs
    end
    return y::Tensor
end

# take a list of potentials, a list of modes to apply to 
# and a list of grids to evaluate
function compute_D(
        potentials::Vec{Function}, # list of potentials
        indices::Vec{<:Vec}, # list of dimensions each potential acts on
        grids::Vec{<:Grid}, # list of grids for each dimension
        beta=1,
    )
        
    y = ones(length.(grids)...)
    for (v, inds) in zip(potentials, indices)
        p = map(Iterators.product(grids[inds]...)) do x
            exp(-1 / 2 * beta * v(x))
        end
        dims = ones(Int, length(grids))
        dims[inds] .= size(p) # specify broadcasting dimensions
        y .*= reshape(p, dims...) # elementwise mult. of potential
    end
    return y::Tensor
end

apply_Q(x, D, E) = apply_A(x .* D, size(E)) ./D .- E .* x

# Example with V(x,y) = x^2 + x^2*y^2 on the domain [-1,1]^2
# For illustrative purposes we represent the potential 
# as sum of a 1 and 2 dimensional potential 
potentials = [x -> x[1] .^ 2, x -> x[1] .^ 2 .* x[2] .^ 2]
indices = [[1], [1, 2]]
grids = [range(-1, 1, 10), range(-1, 1, 10)]

D = compute_D(potentials, indices, grids)
E = apply_A(D, size(D)) ./ D
x = rand(10, 10)
Qx = apply_Q(x, D, E)

\end{lstlisting}

\section{Table of symbols}
\begin{table}[h]
\caption{Table of symbols.
}
\begin{tabular}{ccc}
\hline\hline
                                    & 
Symbol 
                                    & 
Size                                \\ \hline
Number of Cartesian coordinates                                  &
$N$                                                              &
scalar constant                                                  \\
Number of grouped coordinates / subsystems                       &
$n$                                                              &
scalar constant                                                  \\
Number of coordinates of the $i$th subsystem                     &
$L_i$                                                            &
scalar constant                                                  \\
Cartesian coordinates of the full system                         &
$x$                                                              &
$N$-dimensional vector                                           \\
Indices of the subsystems                                        &
$i,j,k,\dots$                                                      &
scalar index                                                     \\
Grouped coordinates of $i$th subsystem                           &
$x_i$                                                            &
scalar variable                                                  \\
Observable function of the combined system                       &
$f(x)$                                                           &
$N$-dimensional function                                         \\ 
Observable function of the coupled system                         &
$\widetilde{f}(x)$                                               &
$N$-dimensional function                                         \\ 
Observable function of the $i$th subsystem                          &
$f_i(x_i)$                                                       &
$L_i$-dimensional function                                       \\ 
Observable function of the $i\sim j$ coupled system                &
$f_{ij}(x_{i},x_j)$                                              &
$(L_i+L_j)$-dimensional function                                         \\ 
Number of subsets in which $x_i$ is discretized                  &
$M_i$                                                            &
scalar constant                                                  \\
Total number of subsets in which $x$ is discretized              &
$M$                                                              &
scalar constant                                                  \\
Indices of the subsets of a discretized coordinate               &
$\alpha,\beta,\gamma,\dots$                                             &
scalar index                                                     \\
Center of the $\alpha$th subset of the discretized coordinate $x_i$ &
$x_i^{\alpha}$                                                      &
scalar variable                                                     \\
\hline\hline
\end{tabular}
\label{tab:tabNotation}
\end{table}

\end{document}